\documentclass[sigconf]{acmart}

\renewcommand\footnotetextcopyrightpermission[1]{} % removes footnote with conference information in first column
\AtBeginDocument{%
  \providecommand\BibTeX{{%
    \normalfont B\kern-0.5em{\scshape i\kern-0.25em b}\kern-0.8em\TeX}}}

\setcopyright{acmcopyright}
\copyrightyear{2020}
\acmYear{2020}
\acmDOI{10.1145/1122445.1122456}

  \acmConference[WWW '2020]{Companion Proceedings of the 2020 World Wide Web Conference (WWW ’20 Companion)}{April 20--24, 2020}{Taipei, Taiwan}

\usepackage{graphicx}
\usepackage{wrapfig}
\usepackage{listings}
\usepackage{lipsum}
\usepackage{subcaption}
\usepackage{caption}
\begin{document}

\title{Wikigender: A Machine Learning Model to Detect Gender Bias in Wikipedia}

\author{Natalie Bolón Brun}
\affiliation{
\institution{École Polytechnique Fédérale de Lausanne (EPFL)}
\city{Lausanne}
  \state{Switzerland}}
\email{natalie.bolonbrun@epfl.ch}

\author{Sofia Kypraiou}
\affiliation{
\institution{École Polytechnique Fédérale de Lausanne (EPFL)}
\city{Lausanne}
  \state{Switzerland}}
\email{sofia.kypraiou@epfl.ch}

\author{Natalia Gullón Altés}
\affiliation{
\institution{École Polytechnique Fédérale de Lausanne (EPFL)}
\city{Lausanne}
  \state{Switzerland}}
\email{natalia.gullonaltes@epfl.ch}

\author{Irene Petlacalco Barrios}
\affiliation{
\institution{École Polytechnique Fédérale de Lausanne (EPFL)}
\city{Lausanne}
  \state{Switzerland}}
\email{irene.petlacalcobarrios@epfl.ch}

%%
%% The abstract is a short summary of the work to be presented in the  article.

%% State the problem, your approach and solution, and the main contributions of the paper. Include little if any background and motivation. Be factual but comprehensive. The material in the abstract should not be repeated later word for word in the paper.

\begin{abstract}
  The way Wikipedia's contributors think can influence how they describe individuals resulting in a bias based on gender. We use a machine learning model to prove that there is a difference in how women and men are portrayed on Wikipedia. Additionally, we use the results of the model to obtain which words create bias in the overview of the biographies of the English Wikipedia. Using only adjectives as input to the model, we show that the adjectives used to portray women have a higher subjectivity than the ones used to describe men. Extracting topics from the overview using nouns and adjectives as input to the model, we obtain that women are related to family while men are related to business and sports.
\end{abstract}



\maketitle

Wikipedia has become a very popular source of information. By November 2019, the number of entries in the English Wikipedia was above 5M \cite{wikipedia-statistics}, and it is increasing every day at a rate of 500 entries on average.

In previous studies, \citeauthor{DBLP:journals/corr/WagnerGJS15} \cite{DBLP:journals/corr/WagnerGJS15} show how gender bias manifest in Wikipedia in the way women and men are portrayed. In a different study, \citeauthor{DBLP:journals/corr/Graells-Garrido15} \cite{DBLP:journals/corr/Graells-Garrido15} show that women biographies are more likely to contain sex-related content. Along with these studies, several other researchers have analyzed topic-related bias in the way women are portrayed, but we can also take a look from a linguistic perspective.

Linguistic bias is defined as a systematic asymmetry in word choice that reflects the social-category cognition that is applied to the described group or individual(s) \cite{linguistic-bias}. We want to analyze how men and women are portrayed and, more specifically, the adjectives used to describe them to spot a possible bias from a linguistic perspective. To do so, we will use the overview of the biographies in the English Wikipedia together with other characteristics of the people we are analyzing.

The overview section of the biographies is the first information to encounter by the reader. According to Wikipedia, it should be written with a neutral point of view and contain a summary of the most relevant content \cite{wiki:Lead_paragraph}. Given this description, we center our study in this section, where the expected non-relevant content is minimal. 

In this work, we model the problem as a prediction task to infer the gender of the person described by using the set of words used in the article. We base our prediction on a logistic regression model that provides interpretable insights on the importance of its features.
The difference is manifested as the presence of different words given the sex of the person being described. This bias is also studied along with different occupations. Finally, we analyze those words that appear as most predictive for each gender and quantify their subjectivity and strength. 

% I think in the paper they get sports an politics and we get sports and business!!

Results show that there is actually a distinction in the usage of words based on gender. As was already shown in \cite{wagner_women_2016}, in terms of topics, women tend to be more related to family and marriage, while men are usually linked to sports and politics topics. Furthermore, results show that women tend to be described using more strongly subjective positive adjectives, while for men, there is a predominance of weakly subjective negative adjectives.

% The perennial question: Should related work be covered near the beginning of the paper or near the end?
% Beginning, if it can be short yet detailed enough, or if it's critical to take a strong defensive stance about previous work right away. In this case Related Work can be either a subsection at the end of the Introduction, or its own Section 2.
% End, if it can be summarized quickly early on (in the Introduction or Preliminaries), or if sufficient comparisons require the technical content of the paper. In this case Related Work should appear just before the Conclusions, possibly in a more general section "Discussion and Related Work".
\section{Related Work}

Gender bias on Wikipedia is a topic that has been widely explored from different perspectives \cite{DBLP:journals/corr/Graells-Garrido15, hill2013wikipedia, wiki-opinion}.

The existence of a gender gap from the editors' perspective was already studied by \citeauthor{hill2013wikipedia} \cite{hill2013wikipedia}, showing a predominant contribution of men who represent more than 70\% of the authors' community. 

In his work "First Women, Second Sex" \cite{DBLP:journals/corr/Graells-Garrido15}, \citeauthor{DBLP:journals/corr/Graells-Garrido15} explore the differences introduced by gender from different perspectives. From the linguistic point of view, they introduce a method to relate topics and gender by exploring the most important n-grams for each gender. Their results show a topical bias given that women are highly related to marriage and family, whether men are linked to sports and politics. These differences also show up in different language editions. 

In \cite{DBLP:journals/corr/WagnerGJS15}, \citeauthor{DBLP:journals/corr/WagnerGJS15} assess the extent to which Wikipedia suffers from potential gender bias. Among others, they explore lexical bias and, by computing log-likelihood ratios, they show that female articles tend to describe romantic relationships and family-related issues much more frequently than male ones in most Wikipedia language editions. 

In "Women through the glass ceiling: gender asymmetries in Wikipedia" \cite{wagner_women_2016}, \citeauthor{wagner_women_2016} analyze different dimensions of the gender gap in Wikipedia. They use Pointwise Mutual Information (PMI) to show that words related to gender, relationships, and family are more prominent for women than men. On the contrary, words associated with men are mainly related to politics and sports. Additionally, for the linguistic bias, they reveal that more abstract terms are used for positive aspects of men's biographies and negatives aspects of women's biographies. This is calculated by computing the ratios of abstract positivity and negativity as the number of positive/negative adjectives over the number of positives/negative words. Apart from topical and linguistic bias, they also show a bias in other dimensions such as notability and structural properties, finding that women in Wikipedia are more notable than men and that structural differences in terms of meta-data and hyperlinks have consequences in information-seeking activities.

The linguistic bias on collaborative crowdsourcing biographies has also been expanded beyond Wikipedia, on the IMDB database by \citeauthor{bias-imdb} \cite{bias-imdb}. She also uses the  Semin and Fiedler’s Linguistic Category Model
(LCM) (Semin and Fiedler 1988) to analyze the biographies. The LCM model classifies terms (like nouns or adjectives) on a scale from abstract to concrete. The more abstract language implies stability
over time and generalizability across situations.

\section{Data}
\subsection{Dataset}
For the following study we use data from the following sources: 
\begin{itemize}
    \item Wikipedia Human Gender Indicators (WHGI) dataset \cite{WHGI}. This dataset contains all the biographic articles from all Wikipedia editions. The version used is that from November the 4th, 2019. From this dataset we extract all biographies appearing in the English Wikipedia together with the corresponding gender. Other data such as the unique identifier of the entry in Wikidata or the occupation of the person are also gathered from this dataset.
    \item Wikidata dataset. It allows us to link the people we want to explore with their corresponding articles in the English Wikipedia. 
    \item Wikipedia dataset. Given the previous steps, we are able to extract the biographies of interest in our study. From them, we will keep only the overviews for the corresponding analysis. 
\end{itemize}

\subsection{Gender and Occupation}
The dataset used is restricted to those entries matched with either male or female gender. In total, nine different gender categories appear along with the whole dataset but those cases not stated as either male or female such as \textit{transgender} or  \textit{non-binary} gender represent less than 1\% of the total number of entries and are not considered in the study. Finally, a total of 1,383,430 articles are used and only 16.58\% of them correspond to female entries. 

\begin{figure}[h]
  \centering
  \includegraphics[width=\linewidth]{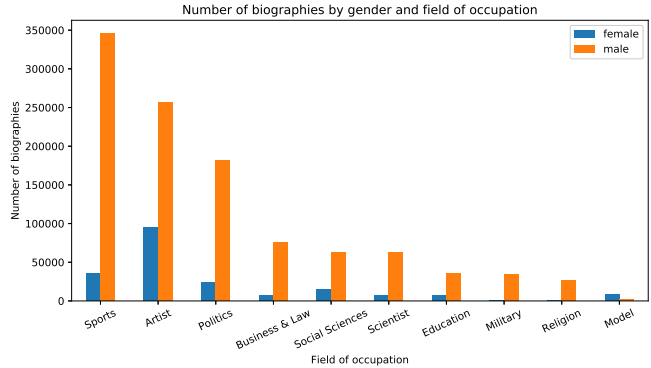}
  \caption{Distribution of biographies per occupation and gender}
  \Description{Distribution of biographies per occupation and gender}
  \label{fig:occu}
\end{figure}

The dataset used is very diverse in terms of occupations. A total of 5,891 different occupations show up with a very different weight in the total representation. We limit the study to the 100 most common occupations since they cover 78.58\% of all the biographies. Moreover, we group them into 10 different fields to extract more meaningful information. As shown in Figure \ref{fig:occu}, the most common occupation is Sports (i.e. footballer, midfielder, etc.) followed by Artist (i.e. actor, singer, painter, etc.) and Politics (i.e. senator, president, etc.). A deeper analysis of the gender by occupation shows again a great disparity with men outnumbering women in all fields except the Model category where the ratio is 5 female entries per each one corresponding to a male.  

\subsection{Subjectivity introduced through the usage of adjectives}
We analyze the subjectivity introduced in the overviews through the usage of adjectives. For this purpose, we make use of the Subjectivity Lexicon version used in \cite{sub_lex}. This allows to determine the degree of subjectivity of the vocabulary and if the given adjectives are usually employed with a positive or negative connotation.

\begin{figure}[h]
  \centering
  \includegraphics[width=\linewidth]{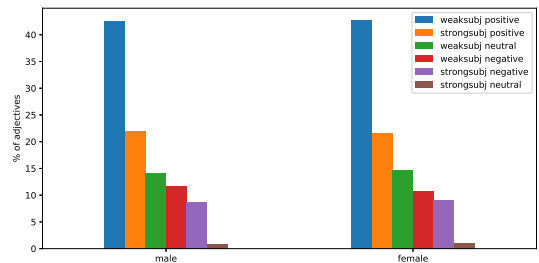}
  \caption{Percentage distribution over the strength and subjectivity of the adjectives}
  \Description{Percentage distribution over the strength and subjectivity of the adjectives}
  \label{fig:adj-subj}
\end{figure}

A first exploration shows that the distribution of strengths and subjectivity of the adjectives is very similar for both genders. In both cases, the majority of adjectives are weakly subjective and positive while those adjectives neutral and strongly subjective are the least present. The final distribution is shown in Figure \ref{fig:adj-subj}. 

In terms of strength, the weakly subjective adjectives account for 68\% of the total adjectives. In terms of subjectivity, the positive adjectives represent 64\% of the adjectives, the negatives represent the 20\% and the neutral ones the 15\%.

If we analyze deeper the most common adjectives for each gender, we can see that three of them (best, high, active) are common for both genders. Moreover, most of them are weakly subjective and have a positive connotation.

\begin{table}[h!]
\centering
\resizebox{0.48\textwidth}{!}{%
\begin{tabular}{|c|c|c|c||c|c|c|}
\hline
 & \multicolumn{3}{c||}{\textbf{MALE}} & \multicolumn{3}{c|}{\textbf{FEMALE}} \\ \cline{2-7} 
\textbf{} & \textbf{adjective} &  \textbf{sentiment} & \textbf{subjectivity strength} & \textbf{adjective} &  \textbf{sentiment} & \textbf{subjectivity strength} \\ \hline
1 & \textbf{active} & positive & \begin{tabular}[c]{@{}c@{}}weak\\ \end{tabular} & \textbf{best} & positive & \begin{tabular}[c]{@{}c@{}}\textbf{strong}\\ \end{tabular} \\ \hline
2 & \textbf{major} &  neutral & \begin{tabular}[c]{@{}c@{}}weak\\ \end{tabular} & \textbf{high} & neutral & \begin{tabular}[c]{@{}c@{}}weak\\ \end{tabular} \\ \hline
3 & \textbf{best} & positive & \begin{tabular}[c]{@{}c@{}}\textbf{strong} \\ \end{tabular} & \textbf{active} & positive & \begin{tabular}[c]{@{}c@{}}weak\\ \end{tabular} \\ \hline
4 & \textbf{high} & neutral & \begin{tabular}[c]{@{}c@{}}weak\\ \end{tabular} & \textbf{popular} & positive & \begin{tabular}[c]{@{}c@{}}weak\\ \end{tabular} \\ \hline
5 & \textbf{famous} & positive & \begin{tabular}[c]{@{}c@{}}weak\\ \end{tabular} & \textbf{long} & negative & \begin{tabular}[c]{@{}c@{}}\textbf{strong}\\ \end{tabular} \\ \hline
\end{tabular}%
}
\caption{Most common adjectives for each gender and their subjectivity and strength.}
\label{tab:adj-sub}
\end{table}

\section{Methodology}\label{methodology}

% bag of words
% encoding
% model
In order to verify the presence of a bias linked to the gender in the overviews of the studied biographies, we develop a model using logistic regression that takes as input a vectorial representation of the text and aims to predict the gender of the person described. 

We first start by encoding the text to obtain a vectorial representation. This process begins by removing stop words such as pronouns. Then we follow with the definition of a vocabulary. This vocabulary is composed by a combination of the top 100 most common words (words being adjectives or adjectives and nouns depending on the model) for each gender from which we obtain a final set of 114 words in the case of adjectives and 132 words when including also nouns. From the vocabulary, we encode the texts in a binary vector in which each entry of the vector represents the presence of a word from the vocabulary in the text. 

The model is build using Logistic Regression. Given the wide diversity of the data in terms of occupation and the imbalanced character of it, we first balance the data per occupation, matching the same number of entries per gender for each occupation. A total of 187,698 entries are then included in the balanced dataset and then split into train and test sets in a proportion 70\%-30\%. 

To verify the robustness and estimate the generalization error, the model is fit and test 50 times. In each case the original dataset is sampled to obtained new balanced version. 

\section{Results}

The bias is studied in two ways: a first model that only uses adjectives and a second model that includes adjectives and nouns. 

\subsection{Model using adjectives}

The model that only uses adjectives achieves an accuracy of 54.6 $\pm$ 0.002\% in the task of determining the gender based on the encoded text. Given that the dataset has been previously balanced, obtaining an accuracy higher than 50\% shows a difference based on gender in the way people are described, i.e. the words used to portray men and women. Apart from discerning the existence of a bias, we can extract from the model those words that are highly indicative of each gender. The adjectives most correlated with female biographies are beautiful, profit, cross, creative and romantic, while the ones most correlated with males are offensive, certain, hard, defensive and diplomatic.

The analysis of the obtained words using the Subjectivity Lexicon \cite{sub_lex} shows that adjectives related to men are weakly subjective and most of them have a negative connotation, whereas the ones related to women are mostly strongly subjective and have a positive connotation (see Table \ref{tab:pred-adj-sub}). Therefore, the overviews portraying females are more likely to contain subjectivity.

\begin{table}[h!]
\centering
\resizebox{0.48\textwidth}{!}{%
\begin{tabular}{|c|c|c|c||c|c|c|}
\hline
 & \multicolumn{3}{c||}{\textbf{MALE}} & \multicolumn{3}{c|}{\textbf{FEMALE}} \\ \cline{2-7} 
\textbf{} & \textbf{adjective} &  \textbf{sentiment} & \textbf{subjectivity strength} & \textbf{adjective} &  \textbf{sentiment} & \textbf{subjectivity strength} \\ \hline
1 & \textbf{offensive} & negative & \begin{tabular}[c]{@{}c@{}}weak\\ \end{tabular} & \textbf{beautiful} & positive & \begin{tabular}[c]{@{}c@{}}\textbf{strong}\\ \end{tabular} \\ \hline
2 & \textbf{certain} &  neutral & \begin{tabular}[c]{@{}c@{}}weak\\ \end{tabular} & \textbf{profit} & positive & \begin{tabular}[c]{@{}c@{}}weak\\ \end{tabular} \\ \hline
3 & \textbf{hard} & negative & \begin{tabular}[c]{@{}c@{}} weak \\ \end{tabular} & \textbf{cross} & negative & \begin{tabular}[c]{@{}c@{}}\textbf{strong}\\ \end{tabular} \\ \hline
4 & \textbf{defensive} & negative & \begin{tabular}[c]{@{}c@{}}weak\\ \end{tabular} & \textbf{creative} & positive & \begin{tabular}[c]{@{}c@{}}\textbf{strong}\\ \end{tabular} \\ \hline
5 & \textbf{diplomatic} & positive & \begin{tabular}[c]{@{}c@{}}weak\\ \end{tabular} & \textbf{romantic} & positive & \begin{tabular}[c]{@{}c@{}}\textbf{strong}\\ \end{tabular} \\ \hline
\end{tabular}%
}
\caption{Most predictive adjectives for each gender and their subjectivity and strength.}
\label{tab:pred-adj-sub}
\end{table}

\subsection{Linguistic bias per field of occupation}

Once we have explored the general linguistic bias, we explore it for each field of occupation to see if there are some occupations with a higher bias than others. We measure the error by randomly balancing the data and computing the accuracy for each model which is fit using data only from the corresponding field of occupation. The fields of Military, Model, and Religion are the only ones where we cannot state that there exists a bias since the accuracy is not significantly above 50\%. These fields are also the ones with fewest data, so this might be the reason behind the high variability of the results. The accuracy for the other fields of occupations is approximately the same, between 55\% and 60\%. The one with the highest bias (i.e. highest accuracy) is Business and Law.

\begin{figure}[h]
  \centering
  \includegraphics[width=0.9\linewidth]{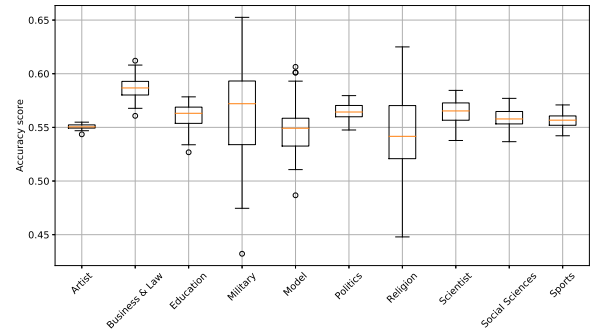}
  \caption{Results of accuracy from the model fit by occupation}
  \Description{Results of accuracy from the model fit by occupation}
  \label{fig:bias-occu}
\end{figure}

\subsection{Model using adjectives and nouns}
A second model using adjectives and nouns is now analyzed to study the effect of the introduction of nouns in the bias. Again, we follow the procedure described in Section \ref{methodology}. Nevertheless, to cope with those words that include references of gender, we substitute them by a neutral form (e.g. actor and actress are substituted by act*). 

The presence of nouns in the text is larger than the one of adjectives which leads to a vocabulary formed mainly by nouns. Among the top 100 most common words (using adjectives and nouns), nouns represent 92\% in the case of females and 91\% for males. The final vocabulary is composed of 132 words and most of them are nouns. 

\begin{table}[h!]
\centering
\resizebox{0.2\textwidth}{!}{%
\begin{tabular}{|c|c|c|}
\hline
 & \textbf{MALE} & \multicolumn{1}{c|}{\textbf{FEMALE}} \\ \hline
1 & footballer & person \\ \hline
2 & war & marriage \\ \hline
3 & officer & model \\ \hline
4 & musician & dancer \\ \hline
5 & football & midfielder \\ \hline
\end{tabular}%
}
\caption{Most predictive words for each gender}
\label{tab:pred-nouns}
\end{table}

In this case, the accuracy achieved by the model rises to 62.9\% $\pm$ 0.002. In this case, the words most correlated with females are person, marriage, model, dancer, and midfielder, while the ones most correlated with males are football, musician, officer, war and footballer. We should mention that "person" includes the words man and woman, but since the words themselves indicate the gender we transform them into a neutral gender one. Words such as spouse and child, which are related to family, also have a positive coefficient which means that they are more predictive for women than men.

\subsection{Words representation}
\begin{figure}[h]
  \centering
  \includegraphics[width=\linewidth]{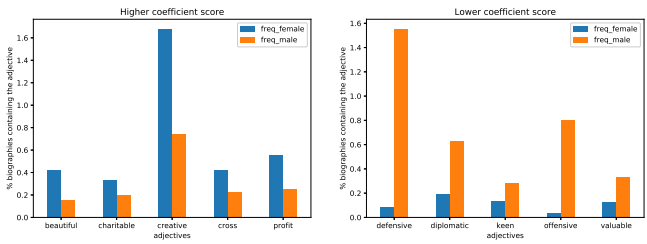}
  \caption{Frequency of appearance of most predictive adjectives along the biographies}
  \Description{Frequency of appearance of most predictive adjectives along the biographies.}
  \label{fig:adj-freq}
\end{figure}

In order to verify the results, we analyze the presence of the most predictive words along with the biographies. As shown in Figure \ref{fig:adj-freq}, those adjectives more correlated to female biographies are more frequent in this group of articles and the same effect occurs in male articles. 

\begin{figure}[h]
  \centering
  \includegraphics[width=\linewidth]{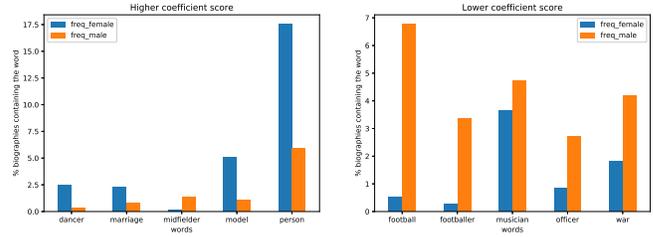}
  \caption{Frequency of appearance of most predictive words along the biographies}
  \Description{Frequency of appearance of most predictive words along the biographies.}
  \label{fig:noun-freq}
\end{figure}

\begin{figure}[h]
  \centering
  \includegraphics[width=\linewidth]{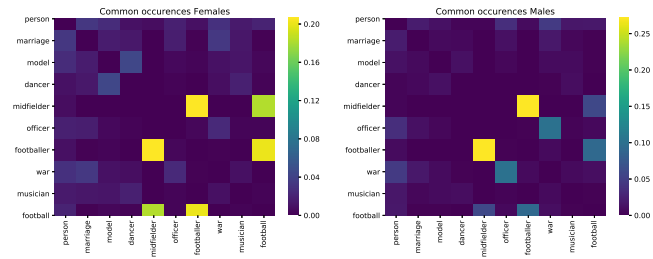}
  \caption{Correlation between words}
  \Description{Correlation between words}
  \label{fig:correlation}
\end{figure}

In the case of the model developed using nouns and adjectives, the majority of the most predictive words for women and for men are more likely to appear in their corresponding overviews. However, the word midfielder is more likely to appear in a man's overview \ref{fig:noun-freq} than in a woman's one although it is among the most predictive words for women. This happens because this word is highly correlated with other words (football and footballer) that are predictive for males, as it can be seen in Figure \ref{fig:correlation}

\subsection{Topic extraction}
Once we know the adjectives and nouns most predictive for males and females, we analyze them using Empath, a tool for analyzing text across lexical categories \cite{fast2016empath}. Using this library, we can extract the categories associated with the words highlighted from the model. We analyze the categories for both the results using the model with only adjectives and the ones using nouns and adjectives.

The results with only adjectives show that women are portrayed as wealthy and men as heroic. Nevertheless, the analysis using both nouns and adjectives results in more insightful results since we are able to extract topics related to them. In this case, we can observe that women are related to family in the first place and other topics related to art (i.e. reading, music), whereas men are mostly related to business and sports.

%% CONCLUSIONS
%% In general a short summarizing paragraph will do, and under no circumstances should the paragraph simply repeat material from the Abstract or Introduction. In some cases it's possible to now make the original claims more concrete, e.g., by referring to quantitative performance results.

%% FUTURE WORK
%% This material is important -- part of the value of a paper is showing how the work sets new research directions. I like bullet lists here. (Actually I like them in general.) A couple of things to keep in mind:
% If you're actively engaged in follow-up work, say so. E.g.: "We are currently extending the algorithm to... blah blah, and preliminary results are encouraging." This statement serves to mark your territory.
% Conversely, be aware that some researchers look to Future Work sections for research topics. My opinion is that there's nothing wrong with that -- consider it a compliment.  

\section{Conclusion and Future Work}

In this work, we presented a different way to measure linguistic and topic bias based on gender in Wikipedia biographies. This new system based on a logistic regression model aims to predict the gender of the person described based only on the appearance of different words in the text. The model also allows us to extract those words that are more relevant for each gender and further analyze them. 

The analysis performed cover from subjectivity introduced by the usage of different adjectives to the extraction of topics based on the most correlated nouns and adjectives for each gender. 

These results show the existence of a difference in the usage of words and topics based on gender. Although these differences may be subtle and could be hidden inside such a great amount of information that Wikipedia constitutes, it is essential to highlight that they should not be normalized, and further steps to limit this issue should be taken. 

Different areas are left as future work after this study. One is the introduction of other features such as year of birth or length of the biographies in order to balance the dataset using propensity score, and therefore eliminate as many confounding factors as possible from the study. Another step to be taken is to extend the study to the whole biographies and not just the overview to see to which extent this bias is present along with the text. In terms of language, extending the study by including other Parts of Speech such as verbs could open a new path to explore. Finally, expanding the work to other languages would be a great way to determine how cultures and their usage of words influence this bias. 

%% Acknowledgements 
%% Don't forget them or you'll have people with hurt feelings. Acknowledge anyone who contributed in any way: through discussions, feedback on drafts, implementation, etc. If in doubt about whether to include someone, include them.
\begin{acks}
We thank Tiziano Piccardi who provided insight and expertise that greatly assisted the research, although he may not agree with all of the interpretations of this paper.
\end{acks}

\bibliographystyle{ACM-Reference-Format}
\bibliography{sample-base}

\end{document}